\documentclass[reprint,superscriptaddress,a4paper,aps,prl,longbibliography]{revtex4-1}
\usepackage{graphicx}
\graphicspath{{./},{/home/user/fig/publi/muon/}}
\usepackage{amsfonts}
\usepackage{amsbsy}
\DeclareMathAlphabet\mathbfcal{OMS}{cmsy}{b}{n}
\begin{document}

\title{Unconventional magnetic order in the conical state of MnSi}

\author{P. Dalmas de R\'eotier}
\affiliation{Universit\'e Grenoble Alpes, INAC-PHELIQS, F-38000 Grenoble, France}
\affiliation{CEA, INAC-PHELIQS, F-38000 Grenoble, France}
\author{A. Maisuradze}
\affiliation{Department of Physics, Tbilisi State University, Chavchavadze 3, 
GE-0128 Tbilisi, Georgia}
\author{A. Yaouanc}
\affiliation{Universit\'e Grenoble Alpes, INAC-PHELIQS, F-38000 Grenoble, France}
\affiliation{CEA, INAC-PHELIQS, F-38000 Grenoble, France}
\author{B. Roessli}
\affiliation{Laboratory for Neutron Scattering and Imaging, Paul Scherrer Institute, 5232 Villigen-PSI, Switzerland}
\author{A. Amato}
\affiliation{Laboratory for Muon-Spin Spectroscopy, Paul Scherrer Institute,
CH-5232 Villigen-PSI, Switzerland}
\author{D. Andreica}
\affiliation{Faculty of Physics, Babes-Bolyai University, 400084 Cluj-Napoca, Romania}
\author{G. Lapertot}
\affiliation{Universit\'e Grenoble Alpes, INAC-PHELIQS, F-38000 Grenoble, France}
\affiliation{CEA, INAC-PHELIQS, F-38000 Grenoble, France}

\date{\today}

\begin{abstract}

In the temperature-magnetic field phase diagram, the binary metallic compound MnSi exhibits three magnetic phases below $T_{\rm c} \approx 29$~K. An unconventional helicoidal phase is observed in zero field. At moderate field intensity a conical phase sets in. Near $T_{\rm c}$, in an intermediate field range, a skyrmion lattice phase appears. Here we show the magnetic structure in the conical phase to strongly depend on the field direction and to deviate substantially from a conventional conical structure.

\end{abstract}

\maketitle

\noindent{\sl Introduction} ---
Since their discovery in the cubic binary metallic compound MnSi (cubic space group P2$_1$3) in 2009 \cite{Muhlbauer09a}, the interest in magnetic skyrmions has been tremendous, mainly explained by their potential applications as information carriers for future magnetic memories \cite{Fert13,Nagaosa13}. In bulk material a  skyrmion lattice phase is observed when an external  magnetic field ${\bf B}_{\rm ext}$ is applied to the system. This phase is surrounded by a conical magnetic phase in which the magnetic moments are believed to have a component parallel to ${\bf B}_{\rm ext}$ and a transverse component akin to the conventional helical structure assigned to the zero-field phase. Interestingly, a skyrmion lattice is described as a coherent superposition of three such helices.
The helical structure is theoretically interpreted by a competition between a dominant ferromagnetic exchange interaction and a weaker Dzyaloshinski-Moriya interaction authorized by the absence of inversion symmetry in the space group P2$_1$3 \cite{Bak80,Nakanishi80}.

Recent results show the magnetic structure of chiral magnets in zero field to be only partly understood. While the magnetic moments in planes perpendicular to the magnetic propagation wavevector ${\bf k}$ are ferromagnetically aligned, their in-plane orientation is not merely given by the scalar product ${\bf k}\cdot{\bf r}$ where ${\bf r}$ defines the position of a magnetic site. An additional phase exists for some position, which was deduced from an analysis of muon spin rotation ($\mu$SR) data in the case of MnSi \cite{Dalmas16} for which ${\bf k}$ is parallel to the [111] crystal axis. In the case where ${\bf k}\parallel [001]$, the symmetry of the crystal also authorizes such a phase \cite{Yaouanc17}.

In the context where the current theories accounting for the wealth of textures found in chiral magnets lead to the conventional helical and conical states, a deeper study of the conical phase is timely. Here we present a detailed refinement of the magnetic structure in the prototypal system MnSi for ${\bf B}_{\rm ext}$ applied along the [111] or [001] crystal axes. According to neutron diffraction, and for high enough fields, we always have ${\bf k} \parallel {\bf B}_{\rm ext}$. For the former direction, we find the additional phase between the moments to be amplified when compared to the zero-field case. For the [001] direction, not only may an additional phase be present but the moments of the helical component are found to rotate in planes not perpendicular to ${\bf B}_{\rm ext}$.

\begin{figure}
{\includegraphics[width=\linewidth]{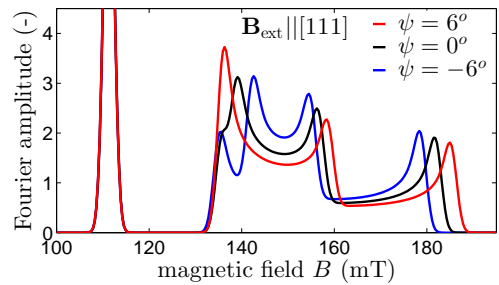}}
\caption{(color online). 
Fourier amplitudes of MnSi $\mu$SR spectra simulated for three values of $\psi$ for ${\bf B}_{\rm ext} \parallel [111]$ with $B_{\rm ext}$ = 200~mT. The angle $\psi$ is defined in Table~\ref{phase}. The other parameters entering the simulations are those obtained from this report except $\lambda_X$ which was set to zero for the sake of clarity.}
\label{sensitivity}
\end{figure}

The experiments were carried with $\mu$SR; for an introduction to the method, see, e.g.\ Ref.~\onlinecite{Yaouanc11}.  With a probe --- the muon --- sitting and measuring the magnetic field at interstitial sites, it is ideally suited for the determination of possible deviations relative to the conventional helices. Figure~\ref{sensitivity} illustrates the sensitivity of $\mu$SR to such small angular deviations for ${\bf B}_{\rm ext}\parallel [111]$.

The measurements were performed relatively close to the magnetic ordering temperature $T_{\rm c} \approx 29$~K, with field magnitudes chosen for a single magnetic domain to be present in the crystal \cite{Grigoriev06}.

The Mn atoms in MnSi occupy $4a$ Wyckoff positions. The coordinates of the four positions, labeled by $\gamma \in \{{\rm I},{\rm II},{\rm III},{\rm IV} \}$, depend on a single parameter $x_{\rm Mn}$ = 0.138. They are $(x_{\rm Mn},x_{\rm Mn},x_{\rm Mn})$, $(\bar{x}_{\rm Mn}+\frac{1}{2},\bar{x}_{\rm Mn},x_{\rm Mn}+\frac{1}{2})$, $(\bar{x}_{\rm Mn},x_{\rm Mn}+\frac{1}{2},\bar{x}_{\rm Mn}+\frac{1}{2})$, and $(x_{\rm Mn}+\frac{1}{2},\bar{x}_{\rm Mn}+\frac{1}{2},\bar{x}_{\rm Mn})$. The lattice parameter is $a_{\rm lat}$ = 4.558~\AA. 

\noindent {\sl Possible magnetic structures in the conical phase} ---
The conical phase is characterized by an incommensurate propagation wavevector $k \approx 0.36$~nm$^{-1}$ \cite{Muhlbauer09a} describing the helical component 
and a second vanishing wavevector responsible for the macroscopic magnetization.  We are left with the determination of the magnetic structure compatible with 
symmetry for the first component.

We shall specify the position of a unit cell by the cubic lattice vector $ {\bf i }$ and that of a Mn atom within a cell by $ {\bf d}_\gamma$. For a magnetic moment at position $ {\bf i + d_\gamma}$ we write 
\begin{eqnarray}
{\bf m}_{i + d_\gamma} & = & {\bf m}_{\rm u} + m_{\rm h} \left (  \cos \alpha_{i, \gamma}\, {\bf a}_{d_\gamma}  - \sin \alpha_{i, \gamma}\, {\bf b}_{d_\gamma} \right )
\label{moment_muon_general}
\end{eqnarray}
setting $\alpha_{i, \gamma}$ = ${\bf k}\cdot ({\bf i}+{\bf d}_\gamma)$. Vector ${\bf m}_{\rm u}$ denotes the uniform component parallel to ${\bf B}_{\rm ext}$ and vectors ${\bf a}_{d_\gamma}$ and ${\bf b}_{d_\gamma}$ together with ${\bf  n}_{d_{\gamma}} \equiv {\bf a}_{d_\gamma} \times {\bf b}_{d_\gamma}$ form a direct orthonormal basis. For ${\bf B}_{\rm ext} \parallel [111]$, the simple solution where ${\bf a}_{d_\gamma}$ and ${\bf b}_{d_\gamma}$ are orthogonal to [111] will be found sufficient, as in zero-field \cite{Dalmas16}. This is not the case for ${\bf B}_{\rm ext} \parallel [001]$. The Euler angles defining ${\bf a}_{d_\gamma}$, ${\bf b}_{d_\gamma}$, and ${\bf  n}_{d_{\gamma}}$ are compiled in Table~\ref{phase} for each sublattice $\gamma$. As an example, the coordinates of ${\bf  n}_{d_{\gamma}}$ are $\left( \cos\varphi_{d_{\gamma}}\sin\theta_{d_{\gamma}},  \sin\varphi_{d_{\gamma}}\sin\theta_{d_{\gamma}}, \cos\theta_{d_{\gamma}} \right )$.
\begin{table}
  \caption{Parameters for the description of the MnSi magnetic structure in the conical phase for ${\bf k} \parallel {\bf B}_{\rm ext} \parallel [111]$ or $[001]$. The results are obtained from representation analysis; see Refs.~\cite{Dalmas16,Yaouanc17}. The table gives the Euler angles $\varphi_{d_\gamma}$, $\theta_{d_\gamma} $, and $\psi_{d_\gamma} $ characterizing the $({\bf a}_{d_\gamma}, {\bf b}_{d_\gamma}, {\bf n}_{d_\gamma}$) basis in the crystal cubic axes.  While the value of $\theta_0$ is fixed ($\cos^2\theta_0 = 1/3$; $\theta_0\approx 54.7^\circ$), the angles $\psi$, $\varphi_2$, $\theta_1$, $\theta_2$ and $\psi_2$ are free parameters of this study.}
\begin{tabular}{c|cccc|cccc}
\hline\hline
& \multicolumn{4}{c|}{${\bf B}_{\rm ext} \parallel [111]$} &  \multicolumn{4}{c}{${\bf B}_{\rm ext} \parallel [001]$} \\ \hline
$\gamma$ & ${\rm I}$ & ${\rm II}$ & ${\rm III}$ & ${\rm IV}$
         & ${\rm I}$ & ${\rm II}$ & ${\rm III}$ & ${\rm IV}$\\ \hline
$\varphi_{d_\gamma} $ & 45$^\circ$ & 45$^\circ$ & 45$^\circ$ & 45$^\circ$
                    & $0$ & 0 & $\varphi_2$ & $\varphi_2$  \\
$\theta_{d_\gamma} $ & $\theta_0$ & $\theta_0$ & $ \theta_0$ & $\theta_0$
                     & $\theta_1$ & $-\theta_1$ & $ -\theta_2$ & $\theta_2$  \\
$\psi_{d_\gamma} $ & $0$ & $\psi$ & $\psi$ & $\psi$
                   & $0$ & 0 & $\psi_2$ & $\psi_2$ \\
\hline\hline
\end{tabular}
\label{phase}
\end{table}

\noindent{\sl The magnetic field at the muon sites} ---
The crystallographic muon site in MnSi has been determined from earlier $\mu$SR measurements \cite{Amato14,Dalmas16}. A density functional theory computation confirms this result \cite{Bonfa15}. The muon sitting at a $4a$ Wyckoff position, four different magnetic sites exist in the cubic unit cell. They are identified with the index $\eta\in\{1,2,3,4\}$. We will denote ${\bf r}_{0,s_\eta}$ the vector distance between a muon position $s_\eta$ and the origin of the cubic lattice.

The local magnetic field ${\bf B}_{{\rm loc},{s_\eta}}$ at position $s_\eta$ comprises ${\bf B}_{\rm ext}$ and the dipolar and contact fields associated with the Mn magnetic moments. Traditionally the dipolar field is split into three terms and accordingly \cite{Schenck95,Dalmas97,Kalvius01}:
\begin{eqnarray}
{\bf B}_{{\rm loc},{s_\eta}}   =  {\bf B}_{\rm ext} + {\bf B}^\prime_{{\rm dip},{s_\eta}} +  {\bf B}_{\rm Lor} +{\bf B}_{\rm dem}  + {\bf B}_{{\rm con},{s_\eta}}.
\label{field_muon_1}
\end{eqnarray}
Here ${\bf B}^\prime_{{\rm dip},{s_\eta}}$ results from the dipolar interaction between the muon magnetic moment and the localized Mn magnetic moments inside the Lorentz sphere and ${\bf B}_{{\rm con},{s_\eta}}$ is the contact field which originates from the polarized conduction electron density at the muon site. Finally, ${\bf B}_{\rm Lor}$ and ${\bf B}_{\rm dem}$ are the macroscopic Lorentz and demagnetization fields. As usual ${\bf B}_{\rm dem} = -N \mu_0 {\bf M}$, where $N$ ($ 0 \le N \le1$) is the demagnetization field factor, $\mu_0$ is  the permeability of free space and ${\bf M}$ = $4\,{\bf m}_{\rm u}/a_{\rm lat}^3$ is the macroscopic magnetization.

Rather than working in direct space, it is convenient to proceed using the reciprocal space \cite{Dalmas16} with \cite{Yaouanc93,Yaouanc93a,Yaouanc11}:
\begin{eqnarray}
& & {\bf B}^\prime_{{\rm dip},{s_\eta}} + {\bf B}_{\rm Lor} + {\bf B}_{{\rm con},{s_\eta}} \label{field_muon_3}\\
& = & \frac{\mu_0}{4\pi} \frac{1}{\sqrt{n_{\rm c}}v_{\rm c}}  
\sum_\gamma \sum_{{\bf q} \in {\rm BZ}} \boldsymbol{J}_{d_\gamma,{\bf q}, s_\eta} {\bf m}_{d_\gamma, {\bf q}} \exp (-i {\bf q} \cdot {\bf r}_{0,s_\eta} ),\nonumber
\end{eqnarray}
where ${\bf m}_{d_\gamma, {\bf q}}$ is the Fourier component of the sublattice magnetic moment. Here $n_{\rm c}$ is the number of unit cells in the crystal under study and $v_{\rm c}$ = $a_{\rm lat}^3$ is their volume. The sum in Eq.~\ref{field_muon_3} is performed over the ${\bf q}$ vectors of the first Brillouin zone (BZ). The unitless tensor ${\boldsymbol J}_{d_\gamma, {\bf q},s_\eta} = {\boldsymbol F}_{d_\gamma, {\bf q},s_\eta} + {\boldsymbol H}_{d_\gamma, {\bf q},s_\eta}$ is the sum of two unitless tensors: ${\boldsymbol H}_{d_\gamma, {\bf q},s_\eta}$ describes the contact interaction parametrized by the quantity $r_\mu H/4\pi$ \cite{Amato14,Dalmas16}, where $r_\mu$ is the number of nearest neighbors used to model the contact interaction, and  ${\boldsymbol F}_{d_\gamma, {\bf q},s_\eta}$ is expressed as a function of a tensor ${\boldsymbol C}_{d_\gamma, {\bf q},s_\eta}$. In recognition of the different nature of the ferro- and antiferromagnetic components of the magnetization the relation between the latter two tensors takes two distinct, albeit related, forms. In terms of the tensor Cartesian components $\alpha\beta$, $F^{\alpha\beta}_{d_\gamma, {\bf q},s_\eta}  = -4\pi\left[ \frac{q^\alpha q^\beta}{q^2} - C^{\alpha\beta}_{d_\gamma, {\bf q},s_\eta}\right]$ for ${\bf q} \ne 0$, and $F_{d_\gamma, {\bf q}=0,s_\eta} = 4\pi C^{\alpha\beta}_{d_\gamma, {\bf q}=0,s_\eta}$ \cite{Yaouanc11}.
The ${\boldsymbol C}_{d_\gamma, {\bf q},s_\eta}$ tensor components are computed following the Ewald summation technique \cite{Ewald21,Born54,Yaouanc93,Yaouanc93a}. This ensures a fast and exact evaluation of the lattice sum which otherwise converges slowly.

The dependence of ${\bf B}_{{\rm loc},{s_\eta}}$ on the magnetic structure is solely described by ${\bf m}_{d_\gamma, {\bf q}}$:
\begin{eqnarray}
{\bf m}_{d_\gamma, {\bf q}} & = &  \sqrt{n_{\rm c}} \left( 
\delta_{{\bf q}, {\bf k} }\,{\tilde {\bf m}}_{d_\gamma,+} +  
\delta_{{\bf q}, -{\bf k} }\,{\tilde {\bf m}}_{d_\gamma,-} +
\delta_{{\bf q}, {\bf 0} }\, {\bf m}_{\rm u} 
\right),\cr
& &
\label{Moment_Fourier_component_1}
\end{eqnarray}
with
\begin{eqnarray}
{\tilde {\bf m}}_{d_\gamma, \pm} = \frac{m_{\rm h}}{2} \left ( {\bf a}_{d_\gamma} \pm  i {\bf b}_{d_\gamma} \right ).
\label{Moment_Fourier_component_2}
\end{eqnarray}
This implies that ${\bf m}_{d_\gamma, {\bf q}}$ vanishes unless ${\bf q} = \pm {\bf k}$ or ${\bf q} = 0$.

Until now we have considered the muon to probe a unique mean magnetic field. However in a typical $\mu$SR experiment millions of muons are implanted in the specimen under study, which localize in different unit cells of the crystal. Therefore the different muons probe the incommensurate magnetic structure for different phases and accordingly we must consider a distribution of local fields rather than a single field. The following field vector distribution is relevant \cite{Dalmas16}: 
\begin{eqnarray}
  D_{\rm v}({\bf B})  & = & \int_{0}^{2\pi}  \delta\left[ {\bf B} 
    - \boldsymbol{\mathfrak{B}}_{{\bf q}=0,{\rm s}_\eta}(0)
    - \boldsymbol{\mathfrak{B}}_{{\bf k},{\rm s}_\eta}(-{\bf k}\cdot {\bf r}_{0,s_\eta}-\zeta) \right. \cr
& &  - \boldsymbol{\mathfrak{B}}_{{\bf -k},{\rm s}_\eta}({\bf k}\cdot {\bf r}_{0,s_\eta}+\zeta) - \left. {\bf B}_{\rm ext} - {\bf B}_{\rm dem} \right] \,{\rm d}\zeta,
\label{distribution_zeta}
\end{eqnarray}
where $\zeta$ is the magnetic structure phase just mentioned. From Eq.~\ref{field_muon_3} we derive $\boldsymbol{\mathfrak{B}}_{{\bf q},s_\eta}(\Psi) = {\mu_0 \over 4 \pi} \frac{1}{\sqrt{n_{\rm c}} v_{\rm c}} \sum_{\gamma} {\boldsymbol J}_{d_\gamma, {\bf q},{\rm s}_\eta} {\bf m}_{{d_\gamma}, {\bf q}} \exp (i \Psi )$.

\noindent {\sl The transverse-field (TF) polarization function} ---
The measurements were performed with a standard TF setup \cite{Yaouanc11} in which the initial muon polarization has a component perpendicular to ${\bf B}_{\rm ext}$ \footnote{At the Swiss Muon Source, as a trade-off between amplitude of spin rotation and counting rate, the angle $\theta_\mu$ between ${\bf B}_{\rm ext}$ and the initial muon beam polarization is lower than 90$^\circ$. While it is known to be around 40 to 50$^\circ$, its actual value varies from a setting to another due to hysteresis in the electromagnet. We take it as a free parameter.}. Conventionally the direction of ${\bf B}_{\rm ext}$ defines the Cartesian $Z$ axis of the spectrometer. The positrons resulting from the anisotropic decay of the muons are detected in counters set along the $X$ axis, perpendicular to $Z$. We denote as $\theta_\mu$ and $\varphi_\mu$ the polar and azimuthal angles of the muon spin at the instant of implantation. Due to the in-flight precession of the muon spins in ${\bf B}_{\rm ext}$ prior to their implantation, $\varphi_\mu$ depends on $B_{\rm ext}$. The evolution of the projection $S_X(t)$ of the muon spin in the local field is given by the solution of the Larmor equation \footnote{We have (see, e.g.\ Eqs.~3.10--13 of Ref.~\onlinecite{Yaouanc11}):
\begin{eqnarray*}
  & & S_X(t)/S & \cr
  & = & \left\{ \frac{B_X^2}{B^2}+\left(1-\frac{B_X^2}{B^2}\right)\cos(\omega_\mu t)\right\} \cos \varphi_\mu\sin\theta_\mu\cr
  & + & \left\{ \frac{B_X B_Y}{B^2}\left[1-\cos(\omega_\mu t) \right] + \frac{B_Z}{B}\sin(\omega_\mu t)\right\}\sin\varphi_\mu\sin\theta_\mu\cr
  & + & \left\{ \frac{B_X B_Z}{B^2}\left[1-\cos(\omega_\mu t) \right] - \frac{B_Y}{B}\sin(\omega_\mu t)\right\}\cos\theta_\mu.
\end{eqnarray*}
Here $\omega_\mu = \gamma_\mu B$ where $\gamma_\mu= 851.616$~Mrad\,s$^{-1}$\,T$^{-1}$ is the muon gyromagnetic ratio and ${\bf B}$ is the local magnetic field at the muon site.}.
The polarization function associated with muons stopped at position $s_\eta$ is obtained after averaging over $D_{\rm v}({\bf B})$:
\begin{eqnarray}
P_{X,s_\eta}(t) = \int \frac{S_X(t)}{S} D_{\rm v}({\bf B})\, {\rm d}^3{\bf B}.
\label{Polarization_th_2}
\end{eqnarray}
Here $P_{X,s_\eta}(t)$ is written in the spectrometer reference frame, while in the previous sections the magnetic structure was expressed in the crystal reference frame. Obviously, proper geometrical transformations of ${\bf B}$ in the spectrometer frame are required before evaluating Eq.~\ref{Polarization_th_2}.

We still need to include the effect of three physical phenomena. We shall do it phenomenologically. As usual, we account for the longitudinal dynamical relaxation with an extra $\exp(-\lambda_Z t)$ factor in the non oscillating components of $S_X(t)/S$. The oscillations are damped by the field distribution produced by the nuclear spins of the compound. Assuming a Gaussian field distribution with a root-mean-square $\Delta_{\rm N}$, we include an $\exp(-\gamma_\mu^2\Delta_{\rm N}^2 t^2/2)$ factor to the oscillating terms in $S_X(t)/S$. Two damping sources of electronic origin are also possible. The first arises from the imperfection of the magnetic structure \cite{Dalmas16}. However, since our data are recorded relatively close to $T_{\rm c}$, we only need to account for the second overwhelming damping induced by magnetic fluctuations. This is achieved by including an $\exp(-\lambda_X t)$ factor to each of the oscillating terms in $S_X(t)/S$.

We cannot distinguish the contribution of the four $s_\eta$ positions in the crystallographic unit cell of MnSi, hence the measured polarization results from the average $P_X(t) = \langle P_{X,s_\eta}(t) \rangle_{\eta}$.

\noindent{\sl Results} ---
Before proceeding to the analysis of the TF $\mu$SR spectra, we expose a few experimental details. 

The platelet-shaped samples were cut from crystals grown by Czochralski pulling and already used in previous measurements \cite{Yaouanc05,Amato14,Dalmas16}. 
The field was applied perpendicular to the platelets. With this geometry we expect  $N  \lesssim 1$. The measurements were performed at 28.14\,(1)\,K for ${\bf B}_{\rm ext} \parallel [111]$, and 
at 27.70\,(1) and 27.85\,(1)\,K for ${\bf B}_{\rm ext} \parallel [001]$. For each of them, the field was systematically applied at 35~K and the sample subsequently cooled down to the desired temperature.

\begin{figure}[t]
\begin{picture}(255,145)
\put(-5,0)
{\includegraphics[width=0.50\linewidth ]{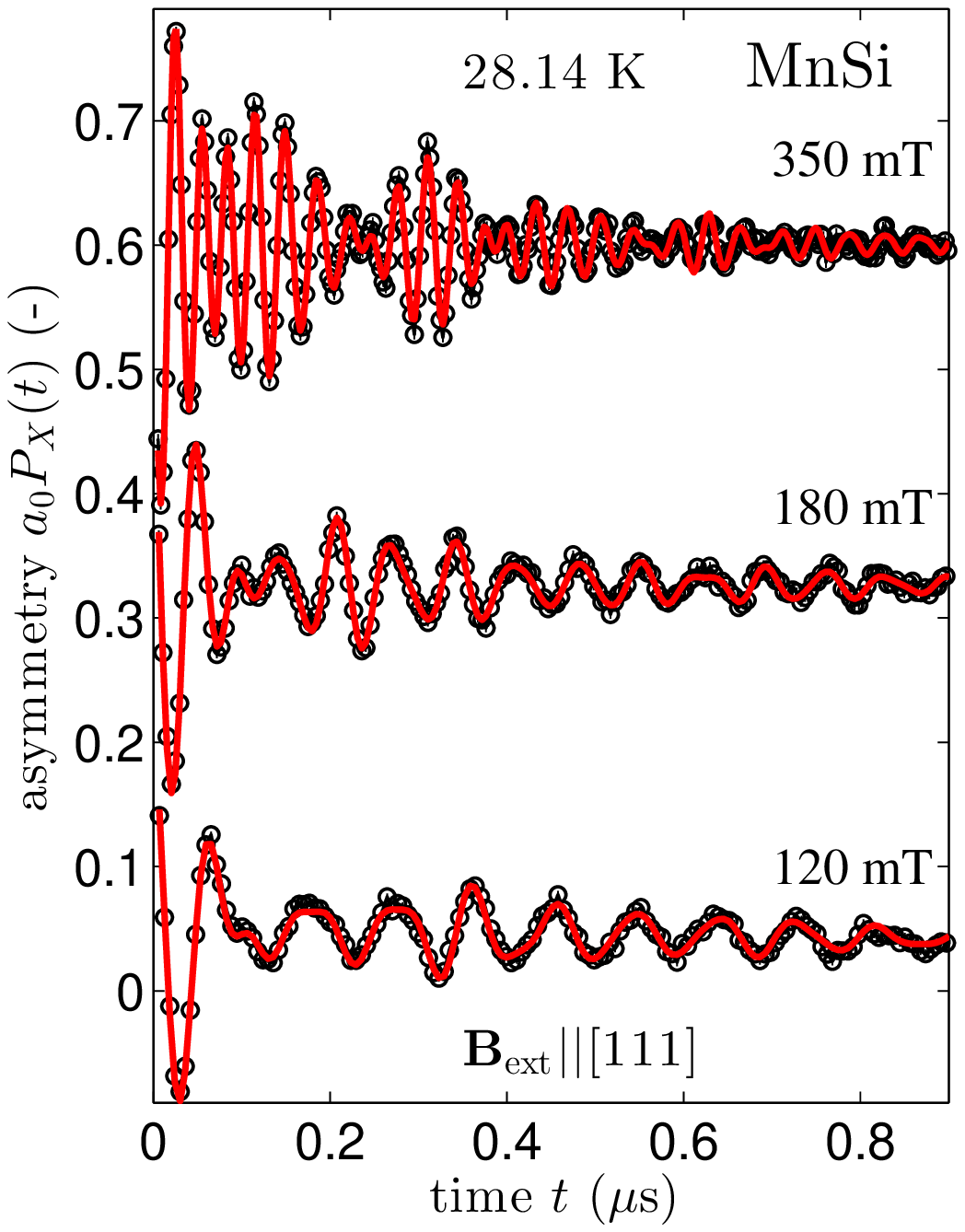}}
\put(0,5){(a)}
\put(123,0)
{\includegraphics[width=0.50\linewidth]{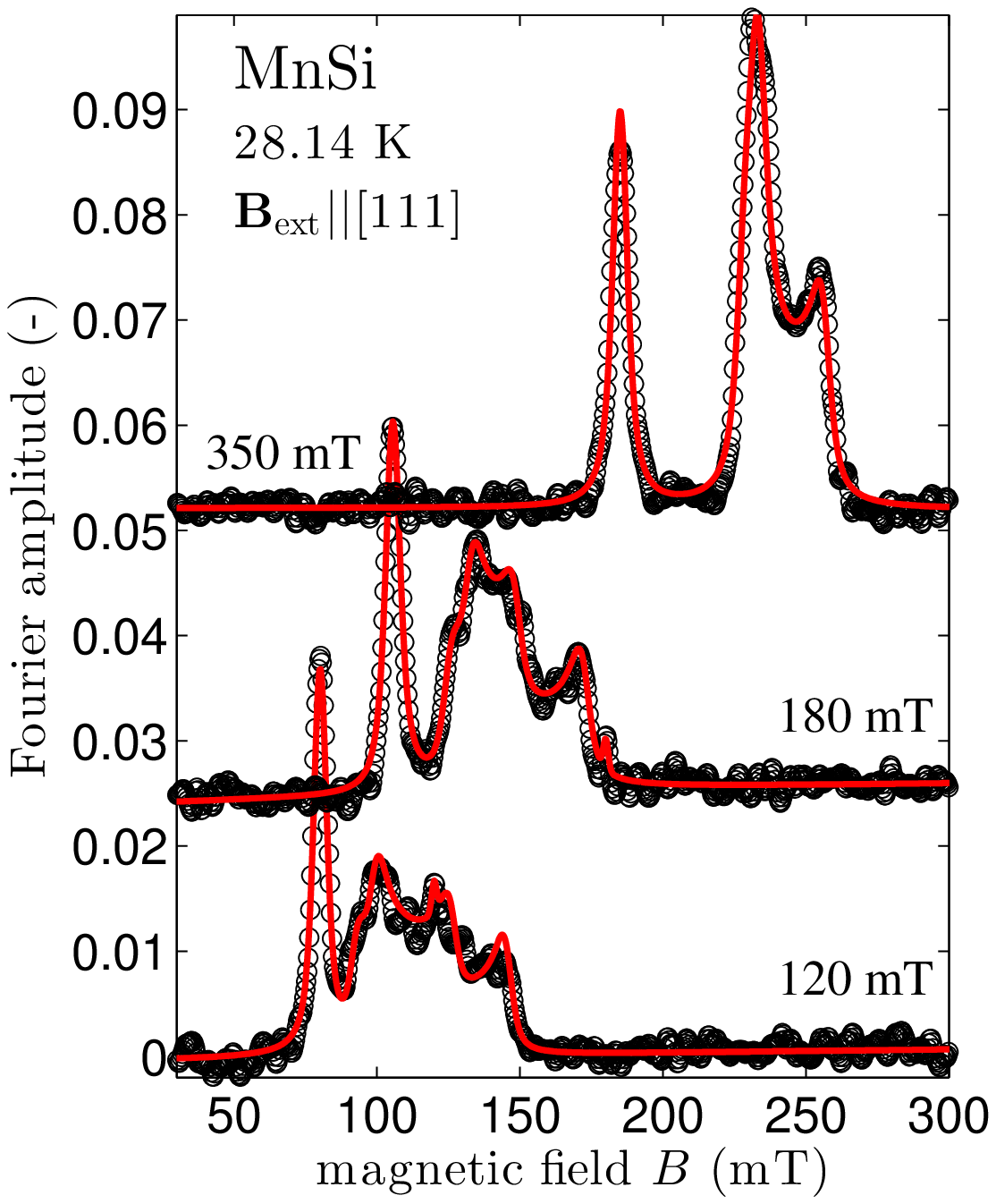}}
\put(128,5){(b)}
\end{picture}
\caption{(color online). 
(a) TF asymmetry spectra recorded at 28.14\,(1)~K in the conical phase of MnSi for different values of $B_{\rm ext}$ with ${\bf B}_{\rm ext} \parallel [111]$. Circles are experimental data while solid lines represent fits as explained in the main text. (b) Real part of the Fourier transforms of the asymmetry spectra precessing component. The solid lines derive from the fits of the asymmetry spectra, i.e.\ we did not fit the Fourier transforms. In line with the large sample size, the contribution of muons stopped in the sample surrounding which can be distinguished for the 180~mT data is negligible. In both panels, the data for consecutive fields are vertically shifted for better visualization.}
\label{Fig_asymmetry_111}
\end{figure}

\begin{figure}[t]
\begin{picture}(255,145)
\put(-5,0)
{\includegraphics[width=0.50\linewidth ]{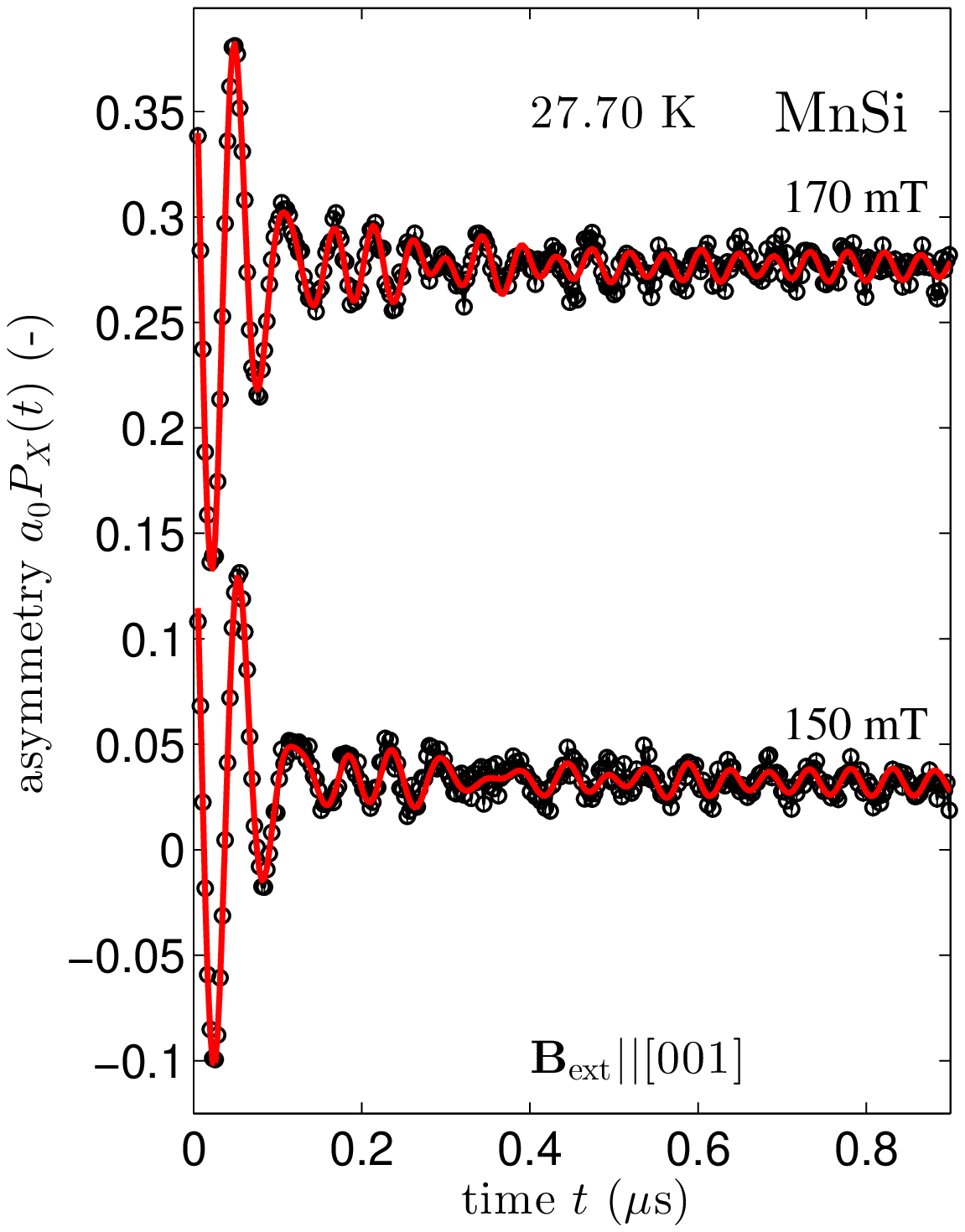}}
\put(0,5){(a)}
\put(123,0)
{\includegraphics[ width=0.50\linewidth]{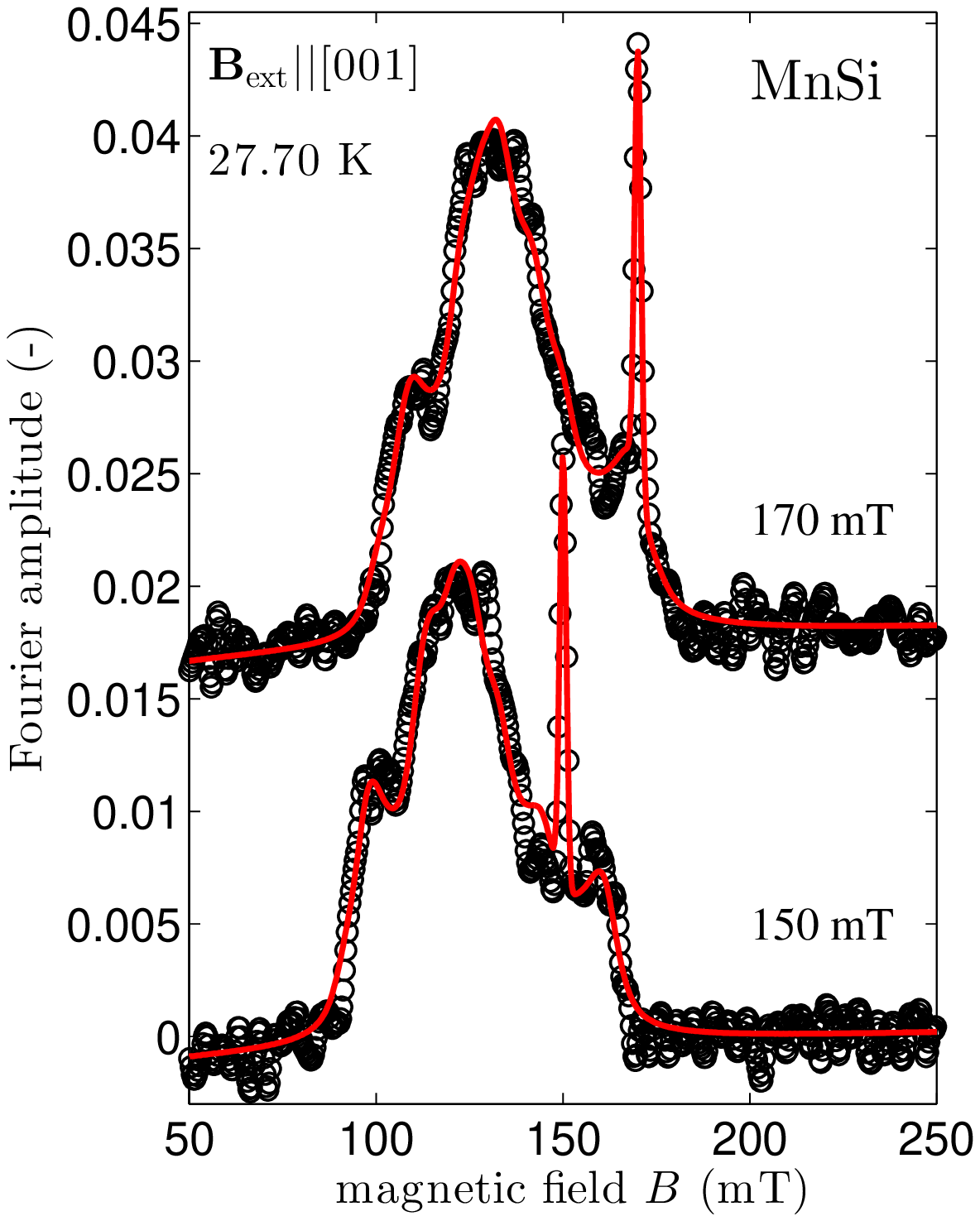}}
\put(128,5){(b)}
\end{picture}
\caption{(color online). 
Same caption as for  Fig.~\ref{Fig_asymmetry_111}, but here the data concern measurements recorded for $T =  27.70\,(1)$~K and ${\bf B}_{\rm ext} \parallel [001]$. The contribution of muons implanted in the sample surroundings is more important than for ${\bf B}_{\rm ext} \parallel [111]$ owing to the smaller sample size.}
\label{Fig_asymmetry_001}
\end{figure}

TF asymmetry spectra are displayed in Figs.~\ref{Fig_asymmetry_111} and \ref{Fig_asymmetry_001}, together with the Fourier transforms of the precessing component.


\begin{figure}[t]
\begin{picture}(255,186)
\put(-5,0)
{\includegraphics[width=0.50\linewidth ]{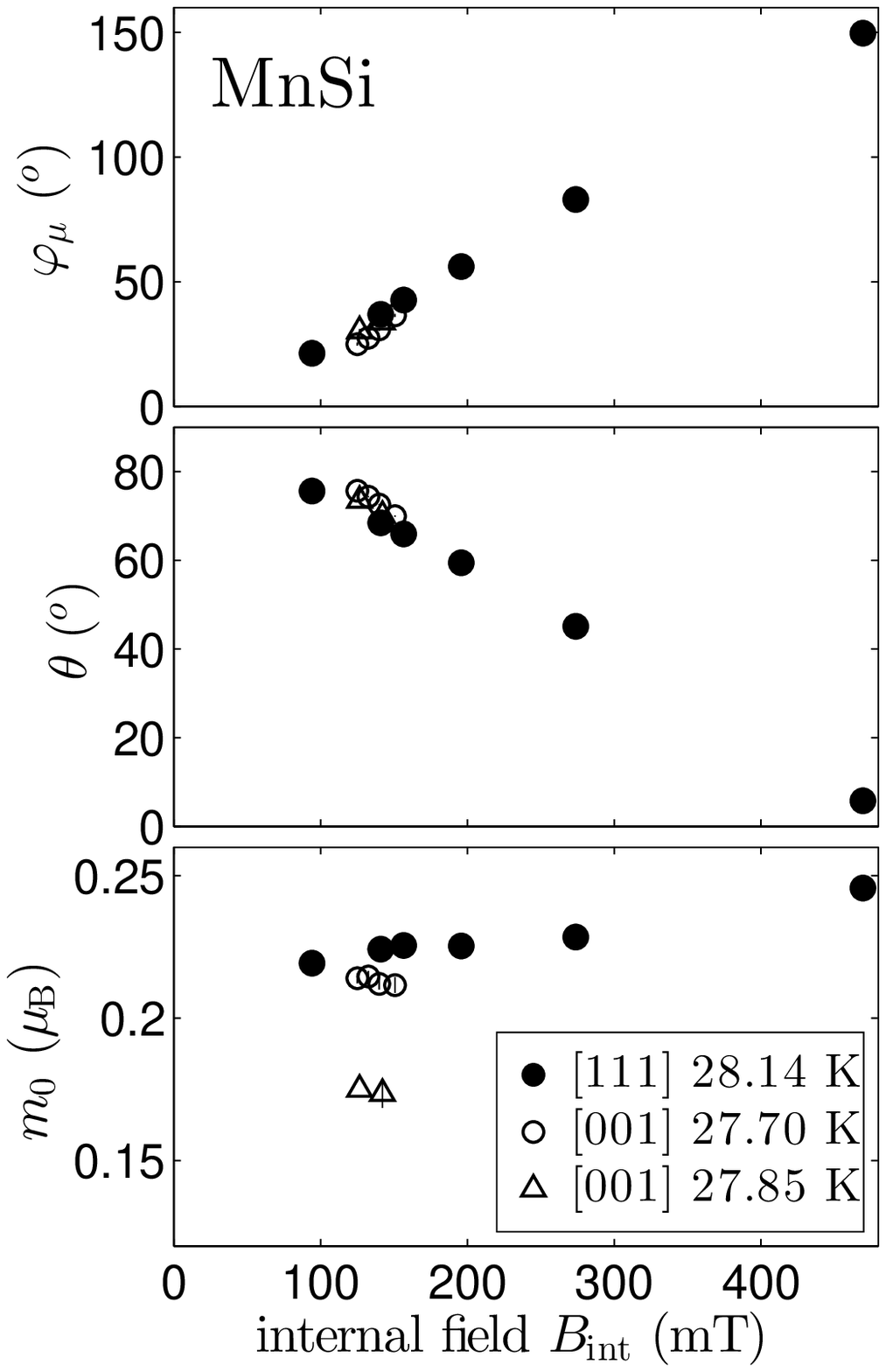}}
\put(0,5){(a)}
\put(120,40)
{\includegraphics[width=0.50\linewidth]{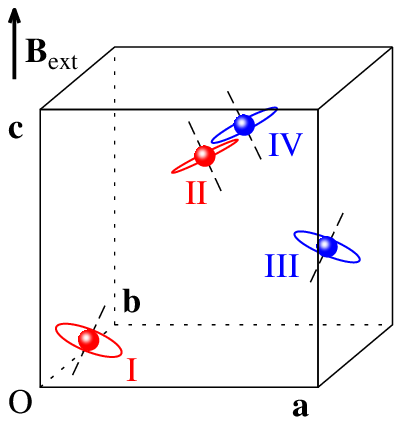}}
\put(125,30){(b)}
\end{picture}
\caption{(color online). 
(a) Parameters characterizing the conical phase of MnSi for ${\bf B}_{\rm ext} \parallel [001]$: $\varphi_\mu$, $\theta \equiv \arctan (m_{\rm h}/m_{\rm u})$, and $m_0 \equiv \left(m^2_{\rm u} + m^2_{\rm h}\right)^{1/2}$. The horizontal scale is the internal field $B_{\rm int} \equiv B_{\rm ext} - \mu_0 N M$. 
The helical components at each of the sublattices is pictured in (b). The normal to the rotation plane is at angle $\pm \theta_1$ or $\pm\theta_2$ from the [001] direction.}
\label{parameters_B}
\end{figure}

For the fits, the following parameters entering in the computation of $P_X(t)$ were fixed to values obtained in previous works \cite{Amato14,Dalmas16}: the muon $4a$ position parameter $x_{\mu^+} = 0.532$, $r_\mu H/ 4 \pi = -1.04$, and $\Delta_{\rm N} = 1.11$~mT. The initial asymmetry $a_0$, a parameter strongly correlated to $\theta_\mu$, was fixed to its theoretical value 0.28. In a first instance the model was fit to individual spectra: some free parameters varied with $B_{\rm ext}$ and others were independent of it. Among the latter parameters, for ${\bf B}_{\rm ext} \parallel [001]$, are $\theta_1$, $\theta_2$, and $\psi_2$. Moreover we also found $\theta_1 \approx -\theta_2$ and $\psi_2\approx 0$ within error bars. In a second step, new fits were performed with field-independent parameters common to all spectra recorded for a given ${\bf B}_{\rm ext}$ direction.
The results are shown as solid lines in Figs.~\ref{Fig_asymmetry_111} and \ref{Fig_asymmetry_001}. Figure~\ref{parameters_B}(a) and Table~\ref{parameters} display the parameters.
\begin{table}
\caption{Model parameters obtained from refinements. While a value common to all field intensities is used for the angles, the relaxation rates slightly depend on $B_{\rm ext}$. Ranges for $\theta_1$ and $\theta_2$ are given. The demagnetization field factor is $N$ = 0.96. }
\begin{tabular}{ccccccc}
\hline\hline
${\bf B}_{\rm ext}$ & $T$        & $\lambda_Z$     & $\lambda_X$     & $\psi$  & $\theta_1 \approx -\theta_2$\\
                    & (K)        & ($\mu$s$^{-1}$) & ($\mu$s$^{-1}$) & ($^\circ$) & ($^\circ$) \\ \hline
$[111]$             & 28.14\,(1) & $\le 0.21$         & $\approx 2.3$     & $-$6.24\,(15) & -- \\
$[001]$             & 27.70\,(1) & $\le 0.1$         & $\approx 3.8$   & --         & [20, 35] \\
$[001]$             & 27.85\,(1) & $\le 0.56$          & $\approx 4.7$   & --         & [20, 35] \\
\hline\hline
\end{tabular}
\label{parameters}
\end{table}

As expected, $\varphi_\mu$ increases linearly with $B_{\rm ext}$; its magnitude is in accord with the spectrometer characteristics \footnote{See the user manual of the GPS spectrometer at https://www.psi.ch/smus/gps.}. This is an independent test of the validity of our analysis.
Considering $\theta$ [Fig.~\ref{parameters_B}(a)], which essentially characterizes the opening angle of the conical structure, its decrease with $B_{\rm ext}$ for the two orientations is anticipated since at high field the magnetic structure tends to be collinear ferromagnetic, i.e.\ $m_{\rm h} \rightarrow 0$. Remarkably, $\theta$ decreases almost linearly with $B_{\rm ext}$. 
Interestingly, the magnetic moment is found smaller for ${\bf B}_{\rm ext} \parallel [001]$ than for [111] despite being measured at a lower temperature.
The angle $\psi$ \footnote{In Ref.~\cite{Dalmas16}, the angular shift between the two orbits was denoted $\phi$. Here it is denoted $\psi$ for consistency with the notation used for ${\bf B}_{\rm ext} \parallel [001]$. $\phi$ and $\psi$ have the same physical significance.} characterizing the shift between the two orbits for ${\bf B}_{\rm ext} \parallel [111]$ is about three times as large as at low temperature in zero field \cite{Dalmas16}. For ${\bf B}_{\rm ext} \parallel [001]$, several solutions with similar confidence parameter $\chi^2$ are found to fit the data equally well for $20^\circ \lesssim \theta_1 \lesssim 35^\circ$, always with $\theta_1\approx -\theta_2$. This substantial value shows that the helical component of the moment is not perpendicular to ${\bf B}_{\rm ext}$ [Fig.~\ref{parameters_B}(b)]. This is the main result for this field orientation. The phase shift $\varphi_2$ resulting from the fit is correlated to $\theta_1$ and $\theta_2$ and can be as large as $\approx 20^\circ$.

\noindent{\sl Discussion} ---
While the equation for a magnetic moment, i.e.\ Eq.~\ref{moment_muon_general}, is generic, the  angles are widely different between the two field orientations.  The difference in the moment values is also noticeable. Both differences reflect the magnetic anisotropy of the system. 
Relevant to this discussion, we previously suggested an additional term to the symmetric and antisymmetric ferromagnetic exchanges and weak anisotropic exchanges to contribute to the  microscopic spin Hamiltonian \cite{Dalmas16}. Thinking in terms of free energy, it would be worthwhile to investigate the effect of anisotropy energy as also proposed for the skyrmion lattice \cite{Everschor11}. 
The two types of deviations from the regular conical structure consistent with representation analysis which are required for the interpretation of experimental data were also derived in a microscopic level theoretical study \cite{Chizhikov12,Chizhikov13}. For ${\bf B}_{\rm ext}\parallel [001]$, the magnetic moment magnitude depends slightly on the Mn position since the (${\bf a}_{d_\gamma}, {\bf b}_{d_\gamma}$) plane is not normal to [001]. An alternative fit was performed replacing Eq.~\ref{moment_muon_general} with ${\bf m}_{i + d_\gamma} = m_{\rm M} {\bf n}_{\rm d_\gamma} + m_{\rm h}\left ( \cos \alpha_{i, \gamma}\,{\bf a}_{d_\gamma} - \sin \alpha_{i, \gamma} {\bf b}_{d_\gamma} \right )$, i.e.\ with a conserved moment. The fit quality is equivalent to that of Fig.~\ref{Fig_asymmetry_001}  and the fit parameters are similar to those of Fig.~\ref{parameters_B}(a) and Table~\ref{parameters}. Therefore we cannot definitively decide between these two related models. 

\noindent{\sl Conclusions} --- A $\mu$SR study of the conical phase in bulk MnSi is reported. A quantitative analysis is performed using symmetry analysis. 
For ${\bf B}_{\rm ext}\parallel [111]$, the phase of the helical component is not solely given by the scalar product ${\bf k}\cdot{\bf r}$. The deviation is enhanced compared to the zero-field case. For ${\bf B}_{\rm ext}\parallel [001]$ the magnetic components associated with the helix rotate in planes which are not perpendicular to ${\bf B}_{\rm ext}$. This information should be helpful for the determination of the microscopic magnetic Hamiltonian.
This work suggests the use of $\mu$SR as a three-dimensional microscopic magnetometry tool, in particular for the helimagnets. Applying this method to the characterization of the skyrmion lattice in MnSi is obviously of great interest.

\begin{acknowledgments}
We acknowledge discussions with V.P.\ Mineev.
This research project has been partially supported by the European Commission under the 7th Framework Programme through the `Research Infrastructures' action of the `Capacities' Programme,  NMI3-II Grant number 283883. Part of this work was performed at the GPS and Dolly spectrometers of the Swiss Muon Source (Paul Scherrer Institute, Villigen, Switzerland).
\end{acknowledgments}

\bibliography{reference}

\end{document}